\documentclass[PRD,twocolumn,showpacs,preprintnumbers,nofootinbib,amsmath,amssymb]{revtex4}

\usepackage{graphicx}
\usepackage{dcolumn}
\usepackage{bm}

\begin{document}

\title{Search with EGRET for a Gamma Ray Line from the Galactic Center}

\author{Anthony R. Pullen, Ranga-Ram Chary, and Marc Kamionkowski}
\affiliation{California Institute of Technology, Mail Code 130-33,
Pasadena, CA 91125}

\date{\today}

\begin{abstract}
We search data from the Energetic Gamma Ray Experiment Telescope
(EGRET) for a gamma-ray line in the energy range 0.1--10 GeV from the 
$10^{\circ} \times 10^{\circ}$ region around the Galactic center.  Our null 
results lead to upper limits to the line flux from the Galactic 
center.  Such lines may have appeared if the dark matter in the
Galactic halo is composed of weakly-interacting massive
particles (WIMPs) in the mass range 0.1--10 GeV.  For a given
dark-matter-halo model, our null
search translates to upper limits to the WIMP two-photon
annihilation cross section as a function of WIMP mass.  We show
that for a toy model in which Majorana WIMPs in this mass range
annihilate only to electron-positron pairs, these upper limits
supersede those derived from measurements of the 511-keV line
and continuum photons from internal bremsstrahlung at the
Galactic center.
\end{abstract}


\maketitle

\section{Introduction} \label{S:intro}

Weakly-interacting massive particles (WIMPs) provide promising candidates for
the dark matter in Galactic halos
\cite{Jungman:1995df,Bergstrom:2000pn,Bertone:2004pz}.  The most
deeply explored WIMP candidate is the neutralino, the lightest
superpartner in many supersymmetric extensions of the standard
model \cite{Haber:1984rc}.
Although the favored mass range for neutralinos
is usually $\gtrsim10$ GeV, there are other WIMP candidates
with masses in the 0.1--10 GeV range.  For example, neutralinos with masses as
low as 6 GeV are plausible if
gaugino unification is not assumed \cite{Bottino:2004qi}.  Neutralinos with
masses as low as 100 MeV are plausible in the next-to-minimal supersymmetric
standard model (NMSSM) \cite{Gunion:2005rw,Ferrer:2006hy}.
Also, scalar and spin-1/2 particles with masses in the MeV range
have been considered \cite{BoehmHooper} to explain the 511-keV
gamma-ray line observed by INTEGRAL
\cite{Knodlseder:2003sv,Jean:2003ci}, a line whose strength, as
explained in Ref.~\cite{BoehmHooper}, has
defied easy explanation from traditional astrophysics.

One way to detect WIMPs is to search for monoenergetic gamma
rays produced by pair annihilation in the
Galactic halo \cite{Bergstrom:1988fp}.  These gamma rays have energies equal to
the WIMP mass $m_\chi$.  Such a line spectrum
could be easily distinguished from the continuum spectrum from
more prosaic gamma-ray sources (e.g., cosmic-ray spallation),
and thus serve as a ``smoking gun'' for dark-matter annihilation.

Since the dark-matter density is highest at the Galactic center,
the flux of WIMP-annihilation photons should be greatest from
that direction.  On the other hand, the continuum background
should also be highest from the Galactic center.
We estimate that for a Navarro-Frenk-White profile~\cite{1996ApJ...462..563N},
the WIMP-annihilation flux from the $10^{\circ} \times 10^{\circ}$ region from
the Galactic center should exceed that from the
Galactic anticenter by a factor $\sim$100, while the flux of cosmic-ray--induced
photons at energies O(GeV) is only about 8 times higher from the Galactic center
than from the Galactic anticenter.  Thus, the Galactic center is the preferred
place to look for a WIMP-annihilation signal.  It is also the location of the
511-keV anomaly that has motivated the consideration of lower-mass WIMPs.

In this paper, we search data from the Energetic Gamma Ray
Experiment Telescope (EGRET) \cite{1993ApJS...86..629T} on the Compton Gamma
Ray Observatory (CGRO) for a gamma-ray line in the energy range
100 MeV to 10 GeV from a $10^{\circ} \times 10^{\circ}$ region
around the Galactic center.  We found no evidence for a
gamma-ray line from the Galactic center in this energy range.  From these null
results, we can bound the
cross section $\langle \sigma v \rangle_{\gamma\gamma}$ for WIMP annihilation
to two photons for WIMPs in this mass range.

The plan of our paper is as follows:  In Section~\ref{S:SD}, we
discuss how EGRET
data are cataloged.  In Section~\ref{S:CGS}, we reconstruct from the EGRET
data the differential flux of photons as a function of energy.  In
Section~\ref{S:DGS}, we fit to the data a model of the flux produced by
cosmic rays and point sources near the Galactic center.  In
Section~\ref{S:APG}, we search for a line excess of photons from
WIMP annihilation.  In Section~\ref{S:VAMR}, we report upper
limits to $\langle \sigma v \rangle_{\gamma\gamma}$ as a function of
$m_{\chi}$ for WIMPs within the mass range of 0.1 GeV to 10 GeV
for a variety of dark-matter-halo models.  In
Section~\ref{S:sigoth}, we show that in a toy model in which the
WIMP annihilates only to electron-positron pairs, this upper
limit is stronger over this mass range than limits derived from
the 511-keV line and from lower-energy continuum gamma rays from
internal bremsstrahlung.

\section{Source of Data} \label{S:SD}

We obtained publicly available data from the CGRO
Science Support Center (COSSC).\footnote{http://cossc.gsfc.nasa.gov/docs/cgro/cossc/egret/}
We used the EGRET photon lists (QVP files), which contain event lists of all photons detected
during a given viewing period.  The data that we used from these files are the
photon's Galactic latitude, Galactic longitude, zenith angle, energy, and
energy uncertainty.  We also required the exposure files, which contain the
detector's effective area multiplied by the viewing time of the detector for a
particular viewing period multiplied by EGRET's 1-sr field of view.  The exposure is provided as a function of latitude,
longitude, and energy range.  We also obtained the counts files, which contain
the number of photons at various spatial coordinates and energy ranges within a
viewing period.  The energy bins, along with their respective energy ranges, are
shown on the COSSC site.

\section{Construction of Gamma Ray Flux} \label{S:CGS}

We begin by constructing the photon differential flux as a
function of energy.  We use data only from a square region on
the sky from $-5^{\circ}$ to $5^{\circ}$ Galactic longitude and
$-5^{\circ}$ to 
$5^{\circ}$ Galactic latitude.  Each viewing period covers a particular region
of the sky, and there were 34 viewing periods for our region of
interest.  These
viewing periods were found using Table 1 in the Third EGRET Catalog
\cite{Hartman:1999fc} and are listed in Table~\ref{T:VPs}.
\begin{table}
\caption{\label{T:VPs} Viewing periods used in analysis.  The more dominant viewing periods are in bold and have an exposure of $>10^6$ cm$^2$ s sr at 150-300 MeV, over our region of interest.}
\begin{ruledtabular}
\begin{tabular}{ccccccc}
\textbf{5.0}&\textbf{7.2}&\textbf{13.1}&\textbf{16.0}&20.0&23.0&\textbf{27.0}\\
35.0&38.0&42.0&43.0&\textbf{209.0}&\textbf{210.0}&\textbf{214.0}\\
219.0&\textbf{223.0}&\textbf{226.0}&229.0&\textbf{229.5}&231.0&\textbf{232.0}\\
\textbf{302.3}&\textbf{323.0}&\textbf{324.0}&\textbf{330.0}&\textbf{332.0}&\textbf{334.0}&\textbf{336.5}\\
339.0&\textbf{421.0}&\textbf{422.0}&\textbf{423.0}&\textbf{423.5}&\textbf{429.0}\\
\end{tabular}
\end{ruledtabular}
\end{table}

The differential photon flux can be determined from the counts
files provided by EGRET, but these provide only counts in 10
energy bins, each with a width comparable to the photon energy in that
bin.  However, we will below search for lines with
energies spanning the full energy range.  This analysis is
performed (as discussed below) by fitting the measured photon
distribution to a continuum plus a line broadened by a Gaussian,
consistent with the instrumental resolution, about each central
line energy.  We therefore work with the EGRET events and exposure files,
which list an energy and effective exposure, respectively, for each photon, and
reconstruct the differential energy flux in 119 energy bins.
Before doing so, however, we first construct the differential
energy flux from the events files with the same 10 bins as in
the EGRET counts files, to be sure that our event-file analysis
recovers the EGRET counts files, the most commonly used EGRET
data product.

\begin{figure}[t!]
\scalebox{.50}{\includegraphics{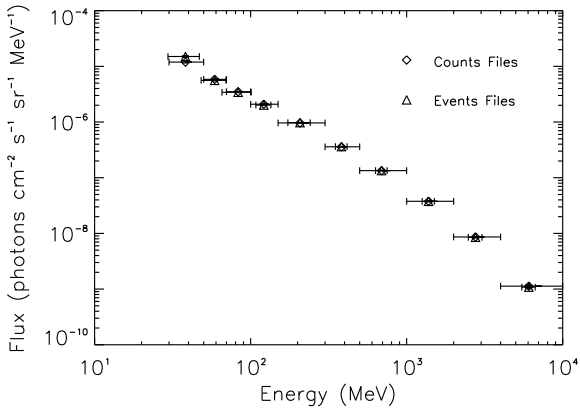}}
\caption{The differential flux within ten energy bins with error bars denoting energy uncertainty for events data and half-bin sizes for counts data.
\label{F:10flux}}
\end{figure}

\begin{figure}[t!]
\scalebox{.50}{\includegraphics{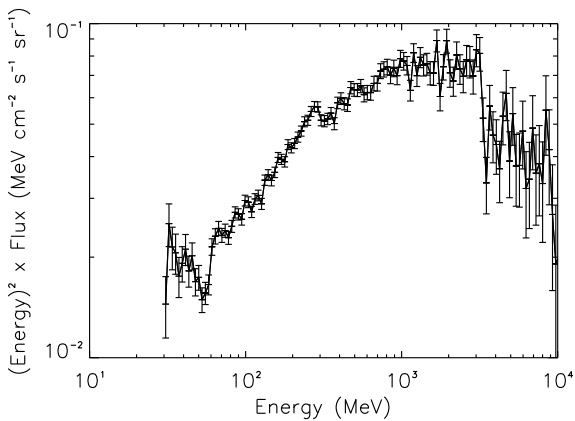}}
\caption{The photon differential flux using 120 energy bins.
\label{F:120flux}}
\end{figure}

We first split the data into the 10 energy bins used by EGRET.  Since the
exposure files record a photon index value of 2.1 for the
photon distribution (a value more-or-less consistent with the
fluxes arrived at in Figs.~\ref{F:10flux} and \ref{F:120flux}),
the average energy $E_{\rm{avg}}$ of photons in an
energy bin [$E_{\rm{min}}$,$E_{\rm{max}}$] is
\begin{equation} \label{E:eavg}
E_{\rm{avg}} = 11 \times \frac{E_{\rm{min}}^{-0.1}-E_{\rm{max}}^{-0.1}}{E_{\rm{min}}^{-1.1}-E_{\rm{max}}^{-1.1}} \mbox{ MeV}.
\end{equation}

Variation of the photon index values over the range [1.7,2.7] only changes
$E_{\rm{avg}}$ by $\sim$1\% for these energy bins and by $\sim$0.01\% for the
119 smaller energy bins.  This variation also only changes the average exposures
by less than $10\%$, which does not affect our final results significantly.
Thus, our assumption of a value of 2.1 for the photon index is a reasonable one.

We calculate the differential flux (photons cm$^{-2}$ s$^{-1}$
sr$^{-1}$ MeV$^{-1}$) from the counts files using
\begin{equation} \label{E:FEi}
F(E_i) = \frac{n(E_i)}{\varepsilon(E_i)\Delta E_i},
\end{equation}
where $E_i$ is the average energy of one of the ten large energy bins, $n(E_i)$ is the number of photons within that energy bin,
$\varepsilon(E_i)$ is the total exposure from the exposure files over the viewing region within that
energy bin, and $\Delta E_i$ is the size of the energy bin.  The quantities
$n(E_i)$ and $\varepsilon(E_i)$ are both summed over all viewing periods and
all positions within the region of interest.  The uncertainty
$\sigma_F(E_i)$ in the flux is
\begin{equation} \label{E:UncF}
\sigma_F(E_i) = \frac{\sqrt{n(E_i)}}{\varepsilon(E_i)\Delta E_i}.
\end{equation}
We assume Gaussian errors in the photon energy.  The energy
uncertainty is just the median of the energy uncertainties of
the individual photons within that energy bin, taken from the
events data.

We then constructed from the events file the photon number
$n(E_i)$ in each counts-file energy bin.  We found that in order
to reproduce the counts data from the events file, we needed to reject
photons with zenith angles greater than $100^{\circ}$ and energy uncertainties
greater than 40\% of the photon energy.  This zenith cut also rejects albedo
gamma rays from the Earth's atmosphere.  
The photon differential fluxes obtained from both the counts
files, and the events files (binned in the same way as the
counts files) are shown in Fig.~\ref{F:10flux}.
We were not able to match the
counts- and events-file photon numbers at the first energy bin to
within 25\%.  However, for reasons discussed below, we discarded
this energy bin (below 0.1 GeV) from our analysis.  

We then proceeded to construct the differential flux
from the events files, applying the same photon cuts, with
narrower bins, to facilitate the analysis in
Section~\ref{S:APG}.
We split the data into 119 energy bins, with each bin ranging in energy from
$E_{\rm{min},i} = 30 \times 1.05^i$ MeV to $E_{\rm{max},i} = 30 \times 1.05^{i+1}$ MeV,
where $i$ ranges from 0 to 118.  To calculate exposures, we
interpolated $\log\lbrack\varepsilon(E_n)\rbrack$ over $\log(E_n)$, where $E_n$
is an average energy for a large energy bin $n$, and $\varepsilon(E_n)$ is the
same exposure for the large bin $n$ used for the ten large bins
earlier.  Fig. 14 of Ref.~\cite{1993ApJS...86..629T} shows that the
exposures do not vary rapidly for energies $\gtrsim0.1$ GeV, and
so this interpolation should be sufficient for our purposes.  The flux
is shown in Fig.~\ref{F:120flux}.  We note that Figs.~\ref{F:10flux}
and~\ref{F:120flux} agree with EGRET's measurement of the diffuse gamma-ray
spectrum in the same region of sky, shown in Fig. 4 of
Ref.~\cite{1997ApJ...481..205H}.  We also note a bump in the differential flux
in Fig.~\ref{F:120flux} at around 3 GeV.  We believe this artifact is due to
the miscalibration of Class B photon events \cite{Willis:1996au}.

\section{Determination of continuum Gamma-Ray Flux} \label{S:DGS}

The line we seek is an excess over a continuum, and we
must therefore model that continuum before we can search for an
excess.  Our aim in this Section is thus to find a simple
functional form that accurately models the continuum over the
resolution scales of the instrument.  A simple linear interpolation
over each space of several energy-resolution elements would be
sufficient, but we instead
consider several astrophysically motivated functional forms,
although the details of the precise astrophysical origin for the
continuum are not important for our search for a line excess.

We were able to find a good fit to the continuum by a linear
combination of three astrophysical sources for the diffuse gamma-ray
background from the Galaxy.  In
the first source, nuclear interactions, cosmic
rays collide with nuclei in interstellar matter to produce neutral pions, which
decay mostly into gamma rays \cite{1993ApJ...416..587B}.  The second process is
bremsstrahlung from cosmic-ray electrons interacting with interstellar matter
~\cite{1993ApJ...416..587B}.  The third, interior-point-source
emission, comes from unresolved point sources within our Galaxy,
such as gamma-ray pulsars \cite{1998ApJ...494..734F}.  We also
considered exterior-point-source emission
\cite{Sreekumar:1997yg} and inverse-Compton scattering of
interstellar radiation from cosmic-ray electrons, but found that
the first three sources listed above were sufficient to model the flux.
Ref.~\cite{1993ApJ...416..587B} gives the differential gamma-ray
production functions for the nuclear and bremsstrahlung contributions.  The
production functions are for the cosmic-ray spectrum in the solar
neighborhood.  We assumed the production functions at the Galactic center are 
proportional to the production functions in the solar neighborhood.

\begin{figure}[t!]
\scalebox{.50}{\includegraphics{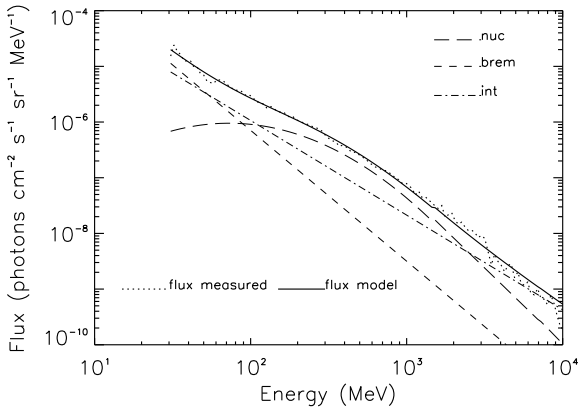}}
\caption{The measured and model gamma-ray flux along with contributions from
nuclear interactions (nuc), bremsstrahlung (brem), and interior point sources (int).}
\label{F:chifit}
\end{figure}

The functional form of the differential flux to which we fitted the data
was $F_{\rm{fit}}(E) = \alpha F_{\rm{nuc}}(E)+\beta F_{\rm{brem}}(E)+\sigma F_{\rm{int}}(E)$,
where $F_{\rm{nuc}}(E)$, $F_{\rm{brem}}(E)$, and $F_{\rm{int}}(E)$ are the
differential photon fluxes from nuclear interactions, bremsstrahlung, and interior point
sources, respectively, and $\alpha$, $\beta$, and $\sigma$ are amplitudes
determined by fitting the data.  The source functions for nuclear
interactions and bremsstrahlung are
\begin{widetext}
\begin{equation} \label{E:qnuc}
F_{\rm{nuc}}(E) = \left\{ \begin{array}{r@{,\quad}l}
2.63\left(\frac{E}{\mathrm{GeV}}\right)^{-2.36}\exp\left[-0.45\left(\ln\left(\frac{E}{\mathrm{GeV}}\right)\right)^2\right] \mbox{ cm$^{-2}$ s$^{-1}$ sr$^{-1}$ GeV$^{-1}$} & 0.01 \mbox{ GeV } < E < 1.5 \mbox{ GeV,} \\
3.3\left(\frac{E}{\mathrm{GeV}}\right)^{-2.71} \mbox{ cm$^{-2}$ s$^{-1}$ sr$^{-1}$ GeV$^{-1}$} & 1.5\mbox{ GeV } < E < 7.0\mbox{ GeV,} \\
4.6\left(\frac{E}{\mathrm{GeV}}\right)^{-2.86} \mbox{ cm$^{-2}$ s$^{-1}$ sr$^{-1}$ GeV$^{-1}$} & E > 7.0\mbox{ GeV,}
\end{array} \right.\, 
\end{equation}
\begin{equation} \label{E:qbrem}
F_{\rm{brem}}(E) = \left\{ \begin{array}{r@{,\quad}l}
0.44\left(\frac{E}{\mathrm{GeV}}\right)^{-2.35} \mbox{ cm$^{-2}$ s$^{-1}$ sr$^{-1}$ GeV$^{-1}$} & 0.01 \mbox{ GeV } < E < 5.0 \mbox{ GeV,} \\
2.1\left(\frac{E}{\mathrm{GeV}}\right)^{-3.3} \mbox{ cm$^{-2}$ s$^{-1}$ sr$^{-1}$ GeV$^{-1}$} & 5.0 \mbox{ GeV } < E < 40 \mbox{ GeV.}
\end{array} \right. \,
\end{equation}
\end{widetext}
We assume interior point sources to be gamma-ray pulsars.  Three pulsars seen
by EGRET were the Crab, Geminga, and Vela pulsars, which have photon indices of
$-2.12$, $-1.42$, and $-1.62$, respectively \cite{1998ApJ...494..734F}.  We approximate the
photon index as having the average value of $-1.7$, so that
\begin{equation} \label{E:qint}
F_{\rm{int}}(E) = \left(\frac{E}{\mathrm{GeV}}\right)^{-1.7} \mbox{ cm$^{-2}$ s$^{-1}$ sr$^{-1}$ GeV$^{-1}$ }.
\end{equation}
The fitted flux [$F_{\rm{fit}}(E_i)$] and the subsequent contributions from each
physical process are shown in Fig.~\ref{F:chifit}.

\section{Analysis of Excess Photons in Gamma Ray Spectrum} \label{S:APG}

\begin{figure}[t!]
\scalebox{.50}{\includegraphics{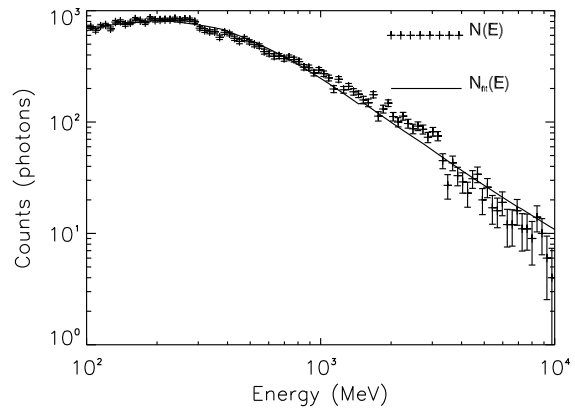}}
\caption{The spectrum of actual counts, $N(E_i)$, and the fitted spectrum, $N_{\rm{fit}}(E_i)$.}
\label{F:difpho}
\end{figure}

We next construct a residual number of counts by subtracting the fitted number
$N_{\rm{fit}}(E_i) = F_{\rm{fit}}(E_i)\varepsilon(E_i)\Delta E_i$
from the observed number $N(E_i)$ of counts.  The counts $N(E_i)$ and
$N_{\rm{fit}}(E_i)$ are displayed in Fig.~\ref{F:difpho}.

We take the residual spectrum to be the upper limit to the number of photons in each
energy bin that could come from WIMP annihilation.  However, to search for the
signal we must take into account the finite energy resolution.  With infinite
energy resolution, the WIMP-annihilation excess would appear as a
monochromatic peak over a smooth background distribution.  However, because of
energy uncertainties, each photon captured by EGRET will appear to have an
energy equal to its true energy plus an error, which we take to be Gaussian.
Thus, monochromatic photons will be spread over neighboring energy bins.
Because our bins are logarithmically spaced, the Gaussian will appear skewed,
but it will still be distinguishable from the background spectrum.

Suppose our true spectrum before measurement consists of a continuum $C(E_i)$
produced by background radiation and an excess $N_p$ of photons with energy
$E_p$.  After measurement, the continuum will change shape but remain smooth,
while the excess will spread out as a Gaussian profile over multiple bins.  The
Gaussian skews negligibly, so we approximate the excess as a standard
Gaussian.  Thus, we model the data $D(E_i)$ as
\begin{equation} \label{E:dgaus}
D(E_i) = C(E_i) + N_pf_p(E_i),
\end{equation}
where $f_p(E_i)$ is a normalized Gaussian of the form,
\begin{equation} \label{E:fgaus}
f_p(E_i) = \frac{\exp\left[-(E_i-E_p)^2/2\sigma^2_{E_p}\right]}{\sum_l\exp\left[-(E_l-E_p)^2/2\sigma^2_{E_p}\right]}.
\end{equation}
In Eq.~(\ref{E:fgaus}), the denominator is summed over all energy bins within
3$\sigma_{E_p}$ of the Gaussian central energy $E_p$.  The energy uncertainty
$\sigma_{E_p}$ at energy $E_p$, is given by
\begin{equation} \label{E:UncE}
\sigma_{E_p} = \frac{E_p}{R(E_p)},
\end{equation}
where $R(E_p)$ is the dimensionless resolution at energy $E_p$.  The fractional full
width at half-maximum (\% FWHM), or $\sqrt{2\ln 2}$ times twice the reciprocal of
the resolution, is shown for various energies in Fig.~20 in
Ref.~\cite{1993ApJS...86..629T}.  From the \% FWHM, we
produce a table of resolution vs. energy, shown in Table~\ref{T:resol}.  We
calculate the
resolution at each energy by interpolating
$\log\lbrack R(E)\rbrack$ over $\log(E)$.  Because the first value for
$R$ given in Table~\ref{T:resol} is for energy $E = 100$ MeV, we cannot
extrapolate $\log(R)$ to lower energies with certainty.  Therefore,
we restrict our analysis to the energy interval 0.1 GeV--10 GeV.
\begin{table}
\caption{\label{T:resol} Dimensionless resolution $R$ of EGRET at various energies.}\begin{ruledtabular}
\begin{tabular}{cc}
Energy (MeV)&$R$\\
\hline
100&9.42\\
200&11.21\\
500&12.39\\
1000&12.08\\
3000&11.49\\
10000&9.07\\
\end{tabular}
\end{ruledtabular}
\end{table}

\begin{figure}[t!]
\scalebox{.50}{\includegraphics{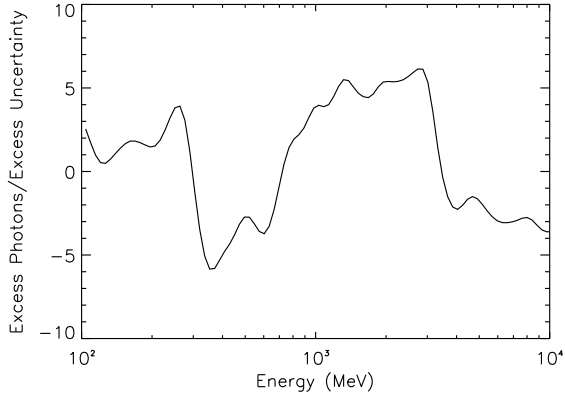}}
\caption{Ratio of excess photons to the excess uncertainty.}
\label{F:numsig}
\end{figure}

The number $N_p(E_i)$ can be deduced at each energy bin in the spectrum by solving
Eq.~(\ref{E:dgaus}) for $N_p$, assuming $D(E_i)$, $C(E_i)$, and $f_p(E_i)$ are
known.  Each $N_p(E_i)$ has an uncertainty,
\begin{equation} \label{E:uncsig}
\sigma_{N_p}(E_i) = \frac{\sqrt{C(E_i)}}{f_p(E_i)},
\end{equation}
due to continuum fluctuations.  Most bins in the
spectrum contain large numbers of
photons.  Therefore, we average $N_p$ using gaussian statistics to calculate
$\overline{N}_p$ and $\sigma_{\overline{N}_p}$, the value and uncertainty of the
excess, for each energy bin $E_p$ greater than 100 MeV.
The resulting ratio of $\overline{N}_p$ to $\sigma_{\overline{N}_p}$ is shown in
Fig.~\ref{F:numsig}.

\begin{figure}[t!]
\scalebox{.50}{\includegraphics{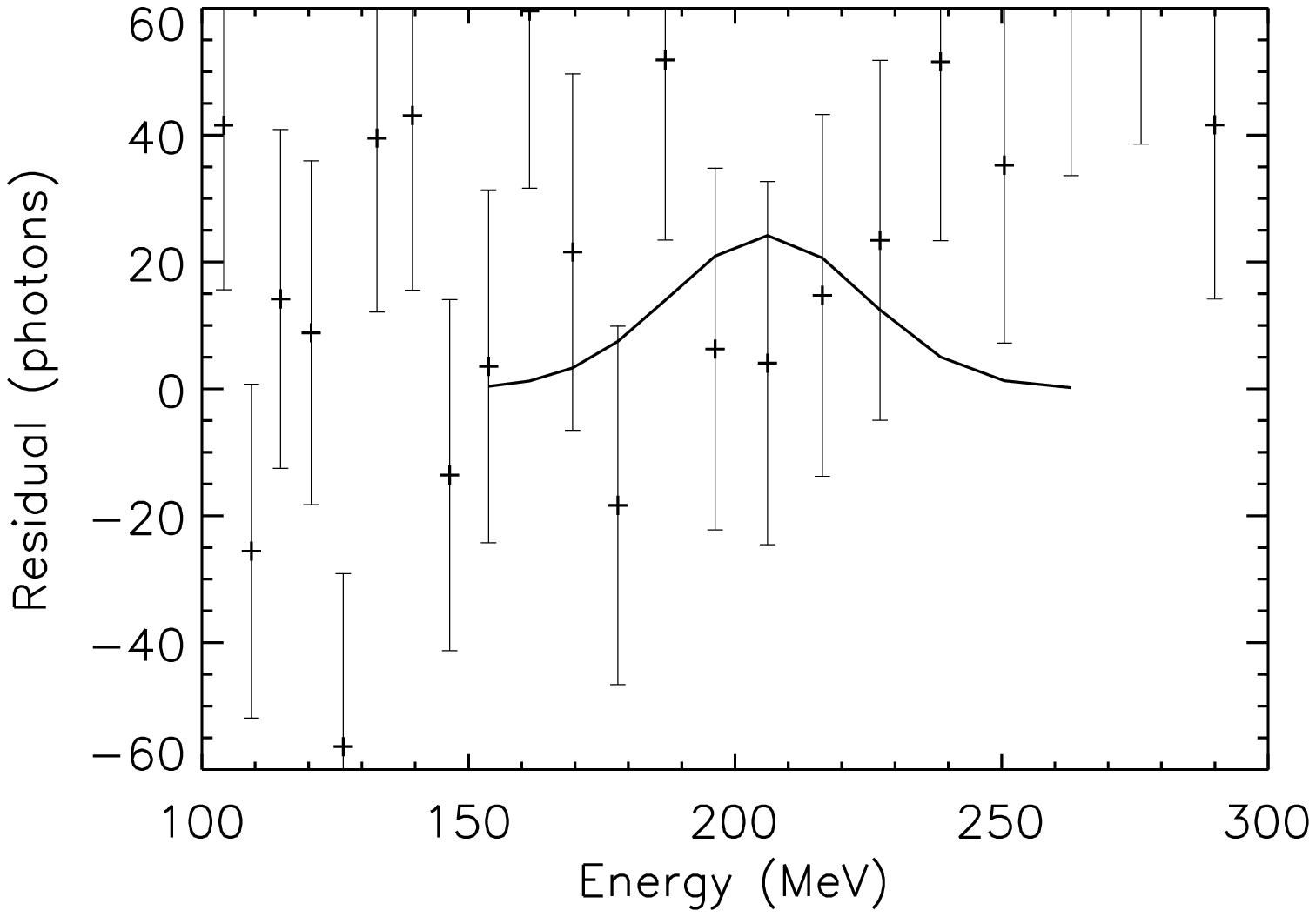}}
\scalebox{.50}{\includegraphics{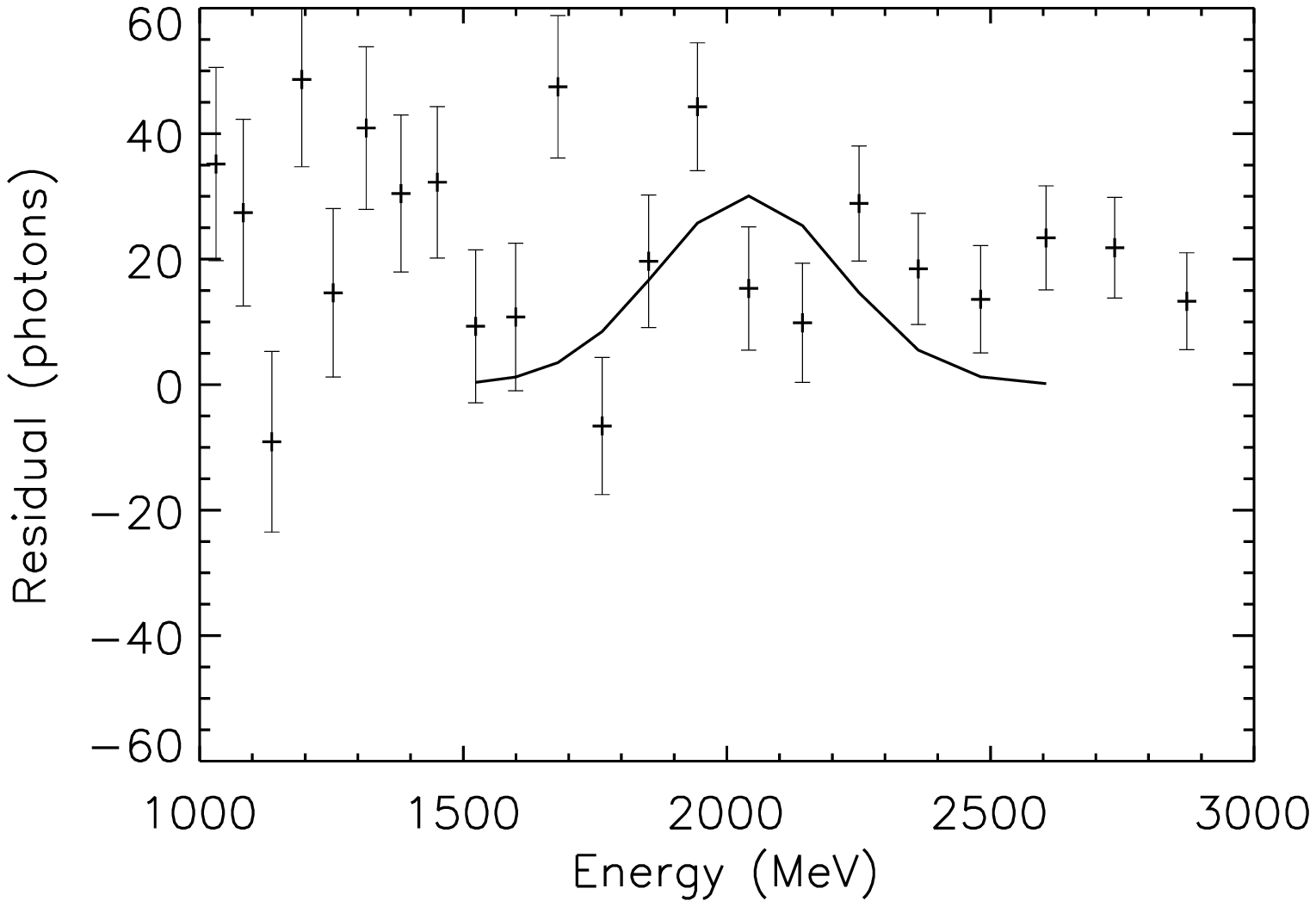}}
\caption{The residual number of counts (crosses) and the expected Gaussian
(solid curve) from a smeared line excess.  The top panel shows the residual and
Gaussian at $E=210$ MeV, while the bottom panel shows the same at $E=2000$ MeV.
The ratio of excess photons to excess uncertainty is high at these energies.
Notice in both panels the residual does not resemble the Gaussian.  For $E=210$
MeV and $E=2000$ MeV, respectively, $\chi^2$ for the Gaussian is 17.3 and 38.0
and $\chi^2$ for the constant excess is 11.0 and 23.9.  Thus, the Gaussian model
is not favored.}
\label{F:gaus}
\end{figure}

Fig.~\ref{F:numsig} does show statistically significant deviations of the data
from our model for the continuum.  To determine if this residual favors the
Gaussian model, we compare $\chi^2$ for a Gaussian model to $\chi^2$ for a
constant-excess model.  We calculate $\chi^2$ for both models
over a $\pm3\sigma_{E_p}$ range centered at the excess center.
The Gaussian is $\overline{N}_pf_p$.  We also compare the residual with a constant excess $N_c$, where
$dN_c/dE$ is constant and $N_c$ is proportional to the energy-bin size.  We
normalize $N_c$ such that the lowest energy bin $3\sigma_{E_p}$ from the Gaussian center has
10 photons.  We compared $\chi^2$ for the excess at energies $E=210$ MeV and $E=2000$ MeV,
two energies that have high excess photons to excess uncertainty ratios (see
Fig.~\ref{F:numsig}).  At both energies we found $\chi^2$ to be smaller for the
constant excess, a simpler model, than for the Gaussian.  Thus, we show that the
residual does not favor the Gaussian model, and we do not attribute any of these
deviations to a WIMP-annihilation line (see Fig.~\ref{F:gaus}).  Rather, it
appears that there is some continuum contribution that our analysis has not
taken into account.

\begin{figure}[t!]
\scalebox{.50}{\includegraphics{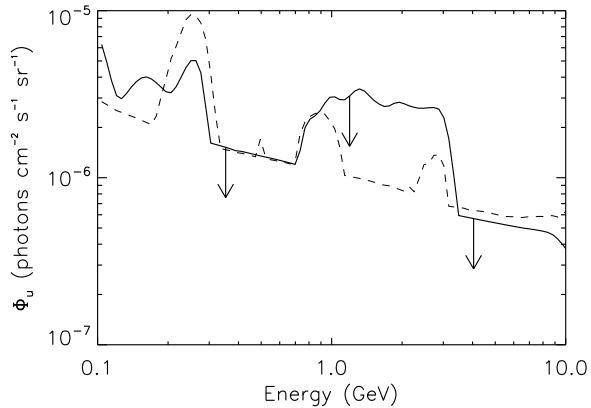}}
\caption{Upper limits to the line flux $\Phi_u$ from the Galactic center.  The solid line is the upper limit derived from the continuum model in Section~\ref{S:DGS}.  The dashed line is the upper limit derived from the sliding window technique.}
\label{F:exflux}
\end{figure}

We therefore use $\overline{N}_p$ to calculate an upper limit
to the line flux.  This line flux is different from the differential flux used
in previous sections in that this flux is not divided by the energy bin size.
Since $\overline{N}_p$ has positive and negative values, we take the $2\sigma$
upper limit to the line flux $\Phi_u(E_p)$ to be
\begin{equation} \label{E:linflux}
\Phi_u(E_p) = \left\{ \begin{array}{r@{,\quad}l}
(\overline{N}_p+2\sigma_{\overline{N}_p})/\varepsilon(E_i) & \overline{N}_p \geq 0\mbox{,} \\
2\sigma_{\overline{N}_p}/\varepsilon(E_i) & \overline{N}_p < 0\mbox{.}
\end{array} \right. \,
\end{equation}
The $2\sigma$ upper limit to the line flux is shown in Fig.~\ref{F:exflux}.

We illustrate the reliability of the upper limit to the line flux by repeating
the analysis in Section~\ref{S:APG} for a sliding-window continuum model.  At
each energy bin $E_i$ we fitted the diffuse flux data within 3 to $9\sigma_{E_i}$ of
$E_i$ to a single power law.  The amplitude and index of the power
law, which varied with energy bin, were then used to construct the background
radiation continuum $C(E_i)$ in Section~\ref{S:APG} needed to search for a line
excess.  No significant excess was found, and an upper limit to the line flux
was determined.  This $2\sigma$ upper limit, shown in Fig. \ref{F:exflux},
agrees quite well with the previous upper limit in Section~\ref{S:APG} except
around 3 GeV, where the previous upper bound is more conservative.  To be
conservative, we chose the upper limit to the line flux from the multi-component
continuum fits, for the rest of our analysis.

\section{Upper Limits to the Annihilation Cross Section} \label{S:VAMR}

If WIMPs comprise the Galactic halo, then the flux of line photons from
WIMP annihilation is (for Majorana WIMPs)
\begin{equation} \label{E:fluxrho}
\Phi(E_{\gamma}=m_{\chi}) = \frac{\langle \sigma v \rangle_{\gamma\gamma}}{4\pi m_{\chi}^2} \int_{l.o.s} \rho_{\chi}^2 \,dl,
\end{equation}
where $\Phi$ is the line flux of photons in units of photons cm$^{-2}$ s$^{-1}$ sr$^{-1}$,
$\langle \sigma v \rangle_{\gamma\gamma}$ is the velocity-averaged
cross section for the WIMP to annihilate to two photons, $m_{\chi}$ is the WIMP
mass (which is equal to the photon energy $E_{\gamma}$), and $\rho_{\chi}$ is
the density profile of the WIMP halo.  The integral is along the
line of sight, and $dl$ is the differential distance along the
line of sight.  The residual in the previous Section gives the
average line-of-sight line flux within a
$10^{\circ}\times 10^{\circ}$ region around the Galactic center.
Therefore, we integrate Eq.~(\ref{E:fluxrho}) over our viewing region to find
the relation between $\langle \sigma v \rangle_{\gamma\gamma}$ and $m_{\chi}$.

The density profile of the WIMP halo must be known in order to integrate
Eq.~(\ref{E:fluxrho}).  The functional form of the halo density profile is
motivated by theory and simulations, with parameters chosen for consistency with
the measured Milky Way rotation curve.  We assume the following parametrization
of the density profile,
\begin{equation} \label{E:profile}
\rho(r) = \rho_0 \frac{(r_0/a)^{\gamma}[1+(r_0/a)^{\alpha}]^{(\beta-\gamma)/\alpha}}{(r/a)^{\gamma}[1+(r/a)^{\alpha}]^{(\beta-\gamma)/\alpha}}.
\end{equation}
Here, $\rho_0$ is the local density of the halo at the Solar System; $r_0$
is the distance from the Solar System to the Galactic center, which we take to
be 8.5 kpc; $a$ is the core radius; and $\alpha$, $\beta$, and $\gamma$ are
parameters that determine the halo model.  Various combinations of $\alpha$,
$\beta$, and $\gamma$ have been used in simulations and are of particular
interest.  We chose to study the Ka and Kb profiles proposed by Kravtsov
et al. \cite{1998ApJ...502...48K}; the NFW profile proposed by Navarro, Frenk, and White
~\cite{1996ApJ...462..563N}; and the modified isothermal profile, or Iso, which is commonly
used.  These profiles are listed in Table~\ref{T:proftype}.  The quantities
$\rho_0$ and $a$ are chosen for each profile so that the profile will account
for the Galactic rotation curve.  These values are taken from
Fig.~5 in Ref.~\cite{Bergstrom:1997fj}.  We insert each of these profiles
into Eq.~(\ref{E:fluxrho}) and
integrate over our viewing region to find the line flux $\Phi$ in terms of
$\langle \sigma v \rangle_{\gamma\gamma}$ and $m_{\chi}$.
\begin{table}
\caption{\label{T:proftype} Parameters for each profile type.}
\begin{ruledtabular}
\begin{tabular}{cccccc}
Profile&$\alpha$&$\beta$&$\gamma$&$\rho_0$ (GeV/cm$^3$)&$a$ (kpc)\\
\hline
Ka&2&3&0.2&0.4&11\\
Kb&2&3&0.4&0.4&12\\
NFW&1&3&1&0.3&25\\
Iso&2&2&0&0.3&4\\
\end{tabular}
\end{ruledtabular}
\end{table}

\begin{figure}[t!]
\scalebox{.50}{\includegraphics{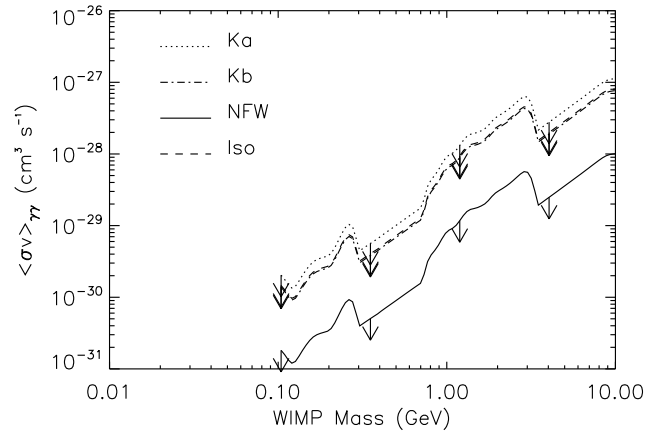}}
\caption{The $2\sigma$ upper limits to the velocity-averaged annihilation cross section $\langle \sigma v \rangle_{\gamma\gamma}$ as a function of WIMP mass for various halo-density profiles.}
\label{F:sigmas}
\end{figure}

The resulting upper limit to the annihilation cross section
$\langle \sigma v \rangle_{\gamma\gamma}$
is shown in Fig.~\ref{F:sigmas} as a function of WIMP mass
$m_{\chi}$ for each halo model listed in Table~\ref{T:proftype}.

\section{Discussion} \label{S:sigoth}

To illustrate the possible utility of this new bound, we
consider a toy model in which WIMPs are Majorana fermions that
couple to electrons via exchange of a scalar boson (the $U$
boson \cite{BoehmFayet,Fayet:2004bw}) of mass $m_U$ (assumed to be much
heavier than both WIMPs and electrons) through the Lagrangian density,
\begin{equation} \label{E:lagrn}
     \mathfrak{L} =
     \frac{C_Uf_{A\,e}}{2m_U^2}\overline{\chi}\gamma_\mu\gamma_5
     \chi\overline{\psi_e}\gamma^\mu\gamma_5\psi_e,
\end{equation}
where $C_U$ and $f_{A\,e}$ are axial couplings of the $U$ boson to
the WIMP field $\chi$ and the electron field $\psi_e$,
respectively.  Annihilation of WIMPs with $O({\rm MeV})$ masses
to electron-positron pairs has been considered as a possible
explanation \cite{BoehmHooper} for the observed flux,
$\Phi_{511} = 9.9^{+4.7}_{-2.1}\times 10^{-4}$ photons cm$^{-2}$
s$^{-1}$~\cite{Jean:2003ci}, of 511-keV photons as measured at the Galactic center
by the SPI camera on the INTEGRAL satellite.
In this scenario, positrons from WIMP
annihilation then annihilate with electrons in the IGM to produce
these 511-keV photons.  The annihilation rate---and 
therefore the cross section for annihilation to
electron-positron pairs and thus the coupling $C_U
f_{A\,e}/m_U^2$---are determined by the flux of 511-keV
photons.  More precisely, the 511-keV flux determines an upper
bound to this annihilation rate, cross section, and coupling,
but we will here suppose the entire 511-keV flux to be from
positrons from WIMP annihilation.

Ref.~\cite{posi} pointed out that if WIMPs annihilate to
electron-positron pairs, they can also undergo annihilation to an electron-positron-photon
three-body final state, a process we refer to as internal bremsstrahlung.  If $\langle \sigma v \rangle_{e^{+}e^{-}}$ is the
cross section for annihilation to electron-positron pairs (as
calculated, e.g., in Refs.~\cite{BoehmFayet,Fayet:2004bw,Goldberg:1983nd}), then the
differential cross section for bremsstrahlung of a photon of
energy $E_\gamma$ is
\begin{equation} \label{E:bsig}
     \frac{d\langle \sigma v \rangle_{\rm{Br}}}{dE_\gamma} =
     \langle \sigma v
     \rangle_{e^{+}e^{-}}\frac{\alpha_e}{\pi}\frac{1}{E_\gamma}
     \left[\ln\left(\frac{s'}{m_e^2}\right)-1\right]\left[1+
     \left(\frac{s'}{s}\right)^2\right], 
\end{equation}
where $s = 4m_\chi^2$, $s' = 4m_\chi(m_\chi-E_\gamma)$, and
$\alpha_e$ is the fine-structure constant.  The
quantity $E_\gamma^2 d\langle \sigma v
\rangle_{\rm{Br}}/dE_\gamma$ increases roughly linearly with
$E_\gamma$ for $E_\gamma < m_\chi$ and peaks at a value (for our
WIMP mass range of 0.1--10 GeV) less than 10\% smaller than the
WIMP mass.  The measured upper limits to the flux were
approximated in Ref.~\cite{posi} $E_\gamma^2 d\Phi_{\rm
Br}/dE_\gamma \lesssim 7\times 10^{-3}$~MeV~cm$^{-2}$~s$^{-1}$~sr$^{-1}$
over the energy range 1--100 MeV.  This flux was averaged over a region on the
sky centered at the Galactic center from $-30^\circ$ to $30^\circ$ Galactic
longitude and $-5^{\circ}$ to $5^{\circ}$ Galactic latitude.  For the purposes
of this
illustrative exercise, we extend this bound up to 10 GeV
(roughly consistent with the line limit we have derived).

Each annihilation to an electron-positron pair produces two 511-keV
photons either directly (7\% of all annihilations) or by producing positronium
and decaying (23.3\% of all annihilations); the rest produce noncontributing
continuum photons \cite{posi,Kinzer:2001ba}.  The resulting flux of 511-keV
photons is (for Majorana particles)
\begin{equation} \label{E:eflux}
     \Phi_{511} = \frac{ \xi \langle \sigma v
     \rangle_{e^{+}e^{-}}}{4\pi m_{\chi}^2} \int
     \rho_{\chi}^2 \,dl\,d\Omega,
\end{equation}
where $\xi=0.303$ is the fraction of positrons that undergo
two-photon annihilation, the $dl$ integral is along the line of
sight and the $d\Omega$ integral is over the SPI camera's field
of view, a $16^\circ$-diameter
circle around the Galactic center.  Likewise, the differential flux of photons from
internal bremsstrahlung is
\begin{equation} \label{E:bflux}
     \frac{d\Phi_{\rm{Br}}}{dE_\gamma} = \frac{d\langle \sigma v
     \rangle_{\rm{Br}}/dE_\gamma}{8\pi m_{\chi}^2\Delta\Omega}
     \int \rho_{\chi}^2 \,dl\,d\Omega,
\end{equation}
where $\Delta\Omega \simeq 0.182$ sr is the solid angle over the $60^\circ$ by
$10^\circ$ Galactic region mentioned earlier.

The two-photon annihilation cross section $\langle
\sigma v \rangle_{\gamma\gamma}$ for the Lagrangian of
Eq.~(\ref{E:lagrn}) is given by
\cite{Rudaz:1989ij}
\begin{equation} \label{E:siggam}
\langle \sigma v \rangle_{\gamma\gamma} = \frac{\alpha_e^2m_\chi^2C_U^2f_{A\,e}^2}{\pi^3m_U^4}\left|I(\xi_e)\right|^2,
\end{equation}
where $\xi_e = m_e^2/m_\chi^2$, $I(\xi_e) =
\frac{1}{2}\left[1+\xi_eJ(\xi_e)\right]$, and $J(\xi_e)$ is
given by
\begin{equation} \label{E:Jxi}
     J(\xi_e) =
     \left(\frac{1}{2}\ln\frac{1+\sqrt{1-\xi_e}}{1-\sqrt{1-\xi_e}}
     - \frac{i\pi}{2}\right)^2,
\end{equation}
for $\xi_e \leq 1$.  For our WIMP mass range 0.1--10 GeV, $\xi_e
\ll 1$ and $I(\xi_e) \simeq 1/2$.  The cross section for annihilation to electron-positron pairs $\langle \sigma v \rangle_{e^{+}e^{-}}$ is given by~\cite{Fayet:2004bw,Goldberg:1983nd}
\begin{equation} \label{E:sigele}
\langle \sigma v \rangle_{e^{+}e^{-}} = \frac{C_U^2f_{A\,e}^2}{2\pi m_U^4}\left[\frac{4}{3}m_\chi^2\overline{v_\chi^2}+m_e^2\right],
\end{equation}
where $\overline{v_\chi^2} = \frac{3}{4}v_c^2$ is the mean-square center-of-mass velocity and
$v_c \simeq 220$ km/s is the WIMP rotation speed, assuming the electron energy $E_e = m_\chi \gg m_e$ and $m_U \gg m_\chi$.
We use Eqs.~(\ref{E:siggam}) and~(\ref{E:sigele}) to derive
upper limits to the coupling $C_Uf_{A\,e}/m_U^2$ appearing in the Lagrangian of
Eq.~(\ref{E:lagrn}).

\begin{figure}[t!]
\scalebox{.50}{\includegraphics{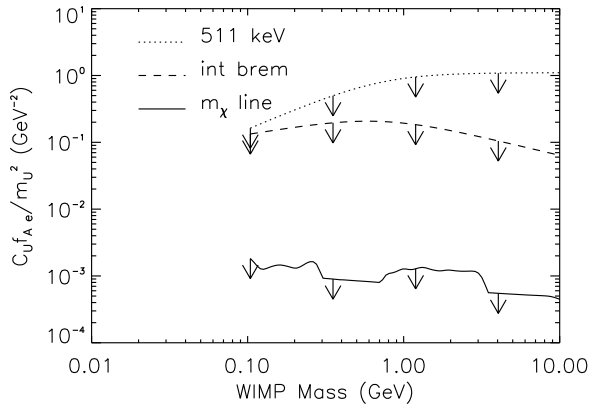}}
\caption{Upper limits to the ratio
     $C_Uf_{A\,e}/m_U^2$ as a function of WIMP mass for the NFW
     halo-density profile.  The limits were calculated from the observed
     511 keV emission, the constraints on internal bremsstrahlung,
     and our derived limit to the line photon flux.}
\label{F:coupoth}
\end{figure}

Fig.~\ref{F:coupoth} shows the upper limit, assuming
an NFW halo-density profile, to the coupling $C_U
f_{A\,e}/m_U^2$ from measurements of the 511-keV line
\cite{BoehmHooper}, the limit to the bremsstrahlung-photon flux
\cite{posi}, and our $2\sigma$ limit to the line-photon flux.  We
see that for the model assumptions and WIMP mass range
considered here, the limit to the two-photon annihilation cross
section derived from our $2\sigma$ limit to the line-photon flux is
the strongest of these three.  At first, this result
may seem surprising, given that the two-photon annihilation
process is higher order in $\alpha_e$, but this suppression is
counteracted by the helicity suppression of the cross section
for annihilation of Majorana fermions to electron-positron
pairs.  Refs. \cite{Beacom:2005qv,Sizun:2006uh} considered also gamma-rays from
in-flight annihilation from e$^+$e$^-$ pairs, but their analysis was
restricted to energies $< 100$ MeV.

Of course, the $2\sigma$ limit to the line-photon flux may not always
provide the best limit to the two-photon annihilation cross
section for every WIMP model.  It may
well be that other models---e.g., those in which the dark-matter
particle is a scalar \cite{Boehm:2006gu}---can produce a ratio of
511-keV photons to line photons large enough to cause the 511-keV limit
to supersede the line photon limit.

\begin{acknowledgments}
We thank S.~Profumo for useful discussions and comments and
J.~Beacom for useful comments on an earlier draft.  ARP was supported
by an NSF Graduate Fellowship.  RC was partially funded under
NASA contract 1407.  MK was supported by DoE DE-FG03-92-ER40701,
NASA NNG05GF69G, and the Gordon and Betty Moore Foundation.
\end{acknowledgments}

\end{document}